\documentclass[twocolumn]{aastex63}

\usepackage{graphicx}  
\usepackage{color}
\usepackage{amsmath}  
\usepackage{verbatim}  
\usepackage{hyperref}
\usepackage[noabbrev,capitalize]{cleveref}  
\usepackage{natbib}
\usepackage[mathscr]{euscript} 
\usepackage{xspace}

\makeatletter
\usepackage{etoolbox}
\patchcmd\H@refstepcounter{\protected@edef}{\protected@xdef}{}{}
\makeatother

\tabletypesize{\scriptsize}  

\newcommand{\sn}{SN Ia\xspace}
\newcommand{\sne}{SNe Ia\xspace}

\newcommand{\h}{\ensuremath{\text{H}_0}\xspace}

\newcommand{\hr}[1][]{Hubble-Lema\^itre residual{#1}\xspace}

\newcommand{\xo}{\ensuremath{x_1}\xspace}
\newcommand{\co}{\ensuremath{c}\xspace}
\newcommand{\gi}{\ensuremath{\gamma_{i}}\xspace}
\newcommand{\gm}{\ensuremath{\gamma_{m}}\xspace}
\newcommand{\gal}{\ensuremath{\gamma_{al}}\xspace}
\newcommand{\gag}{\ensuremath{\gamma_{ag}}\xspace}
\newcommand{\su}{\ensuremath{\sigma_{\mathrm{unexp}}}\xspace}

\newcommand{\unity}{UNITY1.2\xspace}
\newcommand{\salt}{\textsc{SALT2}\xspace}

\newcommand{\pc}[1]{PC\ensuremath{_{#1}}\xspace}
\newcommand{\pcone}{\ensuremath{0.557\xo' - 0.103\co' - 0.535 m' - 0.627 a'_l}\xspace}

\newcommand{\rv}{\ensuremath{R_V}\xspace}
\newcommand{\ha}{\ensuremath{\text{H}\alpha}\xspace}

\newcommand{\un}[1]{~\text{#1}\xspace}  

\begin{document}

\title{Host Galaxy Mass Combined with Local Stellar Age Improve Type Ia Supernovae Distances}
\shorttitle{Mass and Local Age Improve SN~Ia Standardization}

\author[0000-0002-1873-8973]{B. M. Rose}
\affiliation{Department of Physics, Duke University Durham, NC 27708, USA}
\affiliation{Space Telescope Science Institute, 3700 San Martin Drive Baltimore, MD 21218, USA}
\author[0000-0001-5402-4647]{D. Rubin}
\affiliation{Department of Physics and Astronomy, University of Hawai`i at M{\=a}noa, Honolulu, Hawai`i 96822, USA}
\affiliation{E.O. Lawrence Berkeley National Laboratory, 1 Cyclotron Rd., Berkeley, CA 94720, USA}
\author[0000-0002-7756-4440]{L. Strolger}
\affiliation{Space Telescope Science Institute, 3700 San Martin Drive Baltimore, MD 21218, USA}
\author[0000-0003-4069-2817]{P. M. Garnavich}
\affiliation{University of Notre Dame, Center for Astrophysics, Notre Dame, IN 46556, USA}
\collaboration{4}{} 
\correspondingauthor{B. M. Rose}
\email{benjamin.rose@duke.edu}
\shortauthors{Rose, Rubin, Strolger, \& Garnavich}


\date{\today}
\received{October 9, 2020}
\revised{December 2, 2020}
\submitjournal{The Astrophysical Journal}

\begin{abstract}
Type Ia supernovae (SNe~Ia) are standardizable candles, but for over a decade, there has been a debate on how to properly account for their correlations with host galaxy properties.
Using the Bayesian hierarchical model UNITY, we simultaneously fit for the SN~Ia light curve and host galaxy standardization parameters on a set of 103 Sloan Digital Sky Survey II SNe~Ia.  We investigate the influences of host stellar mass, along with both localized ($r<3$~kpc) and host-integrated average stellar ages, derived from stellar population synthesis modeling.
We find that the standardization for the light-curve shape ($\alpha$) is correlated with host galaxy standardization terms ($\gamma_i$) requiring simultaneous fitting. In addition, we find that these correlations themselves are dependent on host galaxy stellar mass that includes a shift in the color term ($\beta$) of $0.8~\mathrm{mag}$, only significant at $1.2\sigma$ due to the small sample.
We find a linear host mass standardization term at the $3.7\sigma$ level, that by itself does not significantly improve the precision of an individual SN~Ia distance. 
However, a standardization that uses both stellar mass and average local stellar age is found to be significant at $>3\sigma$ in the two-dimensional posterior space. In addition,  the unexplained scatter of SNe~Ia absolute magnitude post standardization, is reduced from $0.122^{+0.019}_{-0.018}$ to $0.109\pm0.017$~mag, or $\sim10\%$.
We do not see similar improvements when using global ages.
This combination is consistent with either metallicity or line-of-sight dust affecting the observed luminosity of SNe~Ia.
\end{abstract}

\keywords{\href{http://astrothesaurus.org/uat/1728}{Type Ia supernovae (1728)}; \href{http://astrothesaurus.org/uat/1146}{Observational cosmology (1146)}; \href{http://astrothesaurus.org/uat/394}{Distance indicators (394)}, \href{http://astrothesaurus.org/uat/339}{ Cosmological parameters (339)}}

\section{Introduction}\label{intro}

Type Ia supernovae (\sne) are the runaway thermonuclear burning of a carbon-oxygen white dwarf. Theoretical models suggest that \sne ought to have consistent peak luminosities, i.e. standard candles, and therefore act as precision distance indicators.
However, more detailed observations show that variations in their peak luminosity correlate with other properties (e.g. light-curve shape and color), hence \sne are standardizable candles.
For decades, astronomers have been developing methods to better understand the observed variations in peak luminosity of \sne and improve their use as precision distance indicators. Once basic light curve fitters were developed \citep{Rust1974,Pskovskii1977,Phillips1993,Hamuy1996d,Riess1996,Perlmutter1997,Tripp1998}, \sn became sufficiently accurate cosmological distance indicators to detect the accelerated expansion of the universe \citep{Riess1998,Perlmutter1999}.
 
Many light-curve fitters (e.g., \citealt{Hamuy1996d}, \citealt{Riess1996}, \citealt{Phillips1999}, \citealt{Jha2007}) use a single light-curve shape parameter while separating the observed color variation into a component from intrinsic \sn color and a second to describe the variation due to line-of-sight dust.
Recent work by \citet{Brout2020} reemphasizes the importance of two sources of color variation for precise \sn distances.
Alternatively, \citet{Tripp1998} and \citet{Guy2005, Guy2007}, while still using a single light-curve shape parameter, do not separate the sources of color variation, because this distinction does not appear empirically necessary.

The popular \salt model \citep{Guy2007,Guy2010} is a linearly reduced representation of the diversity of \sn spectral-temporal energy distributions derived from a collection of light curves and spectra \citep{Betoule2014,Mosher2014}. \salt reduces \sn variability down to two parameters. One parameter, $x_1$, captures the ``broader-brighter'' (or Phillips) relationship identified in \citet{Rust1974}, \citet{Pskovskii1977}, and \citet{Phillips1993}. For normal \sne, the distribution of $x_1$ roughly follows a Gaussian distribution with $\mu=0$ and $\sigma=1$. The second parameter, $c$, accounts for color variability both from dust and intrinsic diversity. For typical \sne, $c$ is roughly normally distributed, with a width of $\sim0.1$.

Following the Tripp convention \citep{Tripp1998}, \sn distances can be standardized using the \salt parameters:
\begin{equation}
    \label{eqn:tripp}
    \mu = m_B - \Big(M_B + \alpha \xo + \beta \co
    \Big)
\end{equation}
where $\mu$, $m_B$, $M_B$ are the distance modulus, apparent and absolute B-band magnitude respectively. 
The next two terms are from the \salt model described above.
Both parameters have independent standardization coefficients, $\alpha$ and $\beta$ respectively. Note that for this model $\alpha$ has a minus sign compared to many previous analyses. In this paper, we will quote any external measurements using this convention. 

\citet{Hamuy1995,Hamuy2000} and \citet{Gallagher2005} saw that light-curve shape parameters were correlated with the properties of the host galaxy. Since galaxy properties evolve with redshift, there has been continuous research trying to understand the physical theory --- e.g. multiple \sn channels with dependence on stellar population age and metallicity --- and ultimately the scale of any possible cosmological bias.

However, the correlation seen by \citet{Hamuy2000} became a statistically significant systematic with the works of \citet{Kelly2010}, \citet{Sullivan2010}, and \citet{Lampeitl2010}. It is from these works that we get the so-called ``mass step.'' Though it varies between data sets, the mass step appears to be a $\sim 0.06 \un{mag}$ shift in average standardized peak luminosity of \sn when comparing \sn from low-mass ($\lesssim 10^{10} \un{M}_{\odot}$) and high-mass ($\gtrsim 10^{10} \un{M}_{\odot}$) host galaxies \citep[see][]{Uddin2017a}.
However, the mass step is not ubiquitous, the Dark Energy Survey \citep[DES,][]{DESCollaboration2019} finds no evidence of a mass step \citep{Brout2019}. This does not contradict the evidence seen in other samples since the uncertainties are still relatively large and correlations with selection effects are difficult to quantify \citep{Smith2020}.

The underlying physics responsible for these observed effects remains unclear.
A progenitor metallicity effect was seen prominently in \citet{Moreno-Raya2016a,Moreno-Raya2016b} but not seen in \citet{Kang2020}. Recent star formation rate, an indicator of a prompt explosion ``channel,'' was seen to have a significant effect by \citet{Rigault2013,Rigault2015} using \ha and UV data respectively, however, this is not seen by \citet{Jones2015}.
Similarly, variations in line-of-sight dust properties could produce the observed effects;
\citet{Brout2020} shows that a varying \rv parameter per \sn removes the necessity of the mass step.

A physical explanation, we suspect, should correlate with the local environment (or at least line-of-sight properties) more than the host-integrated average values.
\citet{Rigault2013} used local \ha measurements. Subsequent analyses \citep{Rigault2015,Rigault2018,Jones2015,Jones2018} have, in part, discussed whether these \sn locations are more significant than a random location in the host galaxy. \citet{Jones2018} shows only a marginal ($\lesssim2\sigma$) preference for local environments over a random part of the host galaxy. The works of \citet{Rose2019} and \citet{Kelsey2020} also see no meaningful change between looking at a local stellar population or the entire host galaxy.

\defcitealias{Rose2019}{R19}

The works of \citet{Rigault2018} and \citet{Rose2019} show the largest statistically significant systematics and therefore the greatest likelihood in biasing cosmology. These are not the largest in terms of raw significance, but since these are statistically limited measurements, these highly significant trends with relatively small data sets are unexpected. \citet{Rose2019}, hereafter \citetalias{Rose2019}, see a $4.7\sigma$ correlation with a principal component analysis (PCA) parameter in roughly a factor of 10 less \sn than the $> 5\sigma$ mass step result \citep{Uddin2017a}. 
Similarly the $5.7\sigma$ dependence of \sn standardized luminosity on local specific star-formation rate (lsSFR) of \citet{Rigault2018} used only 40\% more \sn than \citetalias{Rose2019}. 
Interestingly, both results use ``combined''  host galaxy parameters; neither is a simple absolute property, like star formation rate, but rather is scaled by stellar mass.

The Tripp standardization equation (\ref{eqn:tripp}) can be expanded to include host galaxy proprieties. Many apply a step-like function, but a linear correction can be used as well. This results in:
\begin{equation}
    \label{eqn:standardization}
    \mu = m_B - \Big(M_B + \alpha \xo + \beta \co + \sum_{i=1}^N \gamma_i a_i \Big)
\end{equation}
where $\gamma_i$ is the linear standardization term for each host galaxy property (denoted with the place holder $a_i$). Each \sn has its own $\mu$, $m_B$, \xo, \co, and $a_i$ but there is only one value, per data set, of the standardization coefficients ($\alpha$, $\beta$, $\gamma_i$) and the fiducial \sn absolute magnitude ($M_B$).Even though step-like functions are the norm, linear correlations have been seen to adequately fit the data for any exploratory analysis \citep[i.e.][]{Sullivan2010,Ponder2020}.

In this paper, we build upon the work of \citetalias{Rose2019} by simultaneously fitting both the \sn and host galaxy standardization parameters in order to further our understanding of the optimal host galaxy property to use in \sn standardization.
Standardizing \sn with the incorrect host galaxy correlation would produce systematic biases that are large compared to the precision goals of the cosmological surveys of the next decade \citep{Hounsell2018}, however, this is still small compared to the uncertainties of today.
In \cref{sec:method}, we describe our data and analysis method. Followed by a presentation of our results in \cref{sec:results} and a discussion of their implications in \cref{sec:discussion}.

\section{A Simultaneous Standardization Framework}\label{sec:method}

\citetalias{Rose2019} show a systematic in standardized \sn peak luminosity that depended on both host galaxy properties and light curve shape. 
This systematic was derived using PCA on the \sn and host galaxy properties and did not contain any optimization with regard to peak luminosity. The PCA result looked for the largest variance in the four-dimensional light-curve shape, \sn color, host galaxy stellar mass, and age parameter space. This completely ignored the peak magnitude. It was reasonable that one of the eigenvectors in this parameter space could be related to an uncorrected \sn systematic. However, some basic manipulation showed that the PCA eigenvector, though useful, is not the optimal standardization.

To progress, a full simultaneous fit of all standardization parameters is needed to account for all correlations.
In addition to the PCA of \citetalias{Rose2019}, the results of \citet{Roman2018} and \citet{Rigault2018} show a correlation between \sn and host galaxy standardization parameters.
The standard methodology fixes the supernova parameters (i.e. $\alpha$ and $\beta$) and only then searches for \sn-host galaxy correlations in distance residuals. As shown in \citet{Dixon2020}, this method violates assumptions of multi-step linear regression and therefore produces biased results.

In this work, we are replacing the PCA of \citetalias{Rose2019} with a Bayesian hierarchical model (BHM) that can perform the full multidimensional linear regressions while properly handling the uncertainties \citep{Gull1989,Kelly2007} along with the anticipated correlations between light curve parameters and host galaxy properties, originally seen in \citet{Hamuy1995}.
We use the Unified Nonlinear Inference for Type Ia cosmologY \citep[UNITY,][]{Rubin2015}, specifically a more recent version that includes the capability of modeling Tripp-like standardization equations \citep{Tripp1998} with an arbitrary number of standardization parameters \citep[\unity,][]{Rose2020a}. These latest updates to UNITY can be found at \url{https://github.com/rubind/host_unity}.

\subsection{The Data Set}

For our simultaneous standardization, we use the data presented in Table 7 of \citetalias{Rose2019}.
This is a relatively small data set ($N=103$), but it has quality local and global stellar age estimates ($a_l$ and $a_g$ respectively). 
The final data set is a subset of the initial spectroscopic and photometric classified \sn sample from SDSS-II Supernova survey \citep{Sako2008,Campbell2013} with an additional redshift cut, $z < 0.2$, applied.
For now, the \sn data set of \citet{Campbell2013}, along with the host galaxy properties presented in \citetalias{Rose2019} can be used to demonstrate this simultaneous fitting method. Note that the \sn from \citet{Campbell2013} do not have the host mass correction, and only applies a simple one-dimensional redshift-based Malmquist bias correction. 

\unity requires that the data be described with Gaussian uncertainties. Many parameters are already represented this way, but the age estimates used in \citetalias{Rose2019} were numerical representations of the probability distribution and typically non-Gaussian (see \citealt{Rose2019-data} for the full numerical representations).
These ages were estimated by fitting a four parameter delayed-tau star formation history from the Flexible Stellar Population Synthesis \citep[FSPS; ][]{Conroy2009,Conroy2010} code to the observed SDSS \textit{ugriz} photometry.
Since spectral energy distribution based ages are not more than a course estimation --- with typical uncertainties of $\sim 0.3 \un{dex}$ --- a single Gaussian can be fit to the non-Gaussian probability distributions with minimal loss of information. It was this Gaussian representation that was reported in Table 7 and used in the original PCA. 
There is still enough information to observe a systematic caused by explosion mechanism (seen in a difference in young verse old progenitors) or from a more continuous build-up of metals.

\begin{figure}
    \centering
    \includegraphics[width=.99\columnwidth]{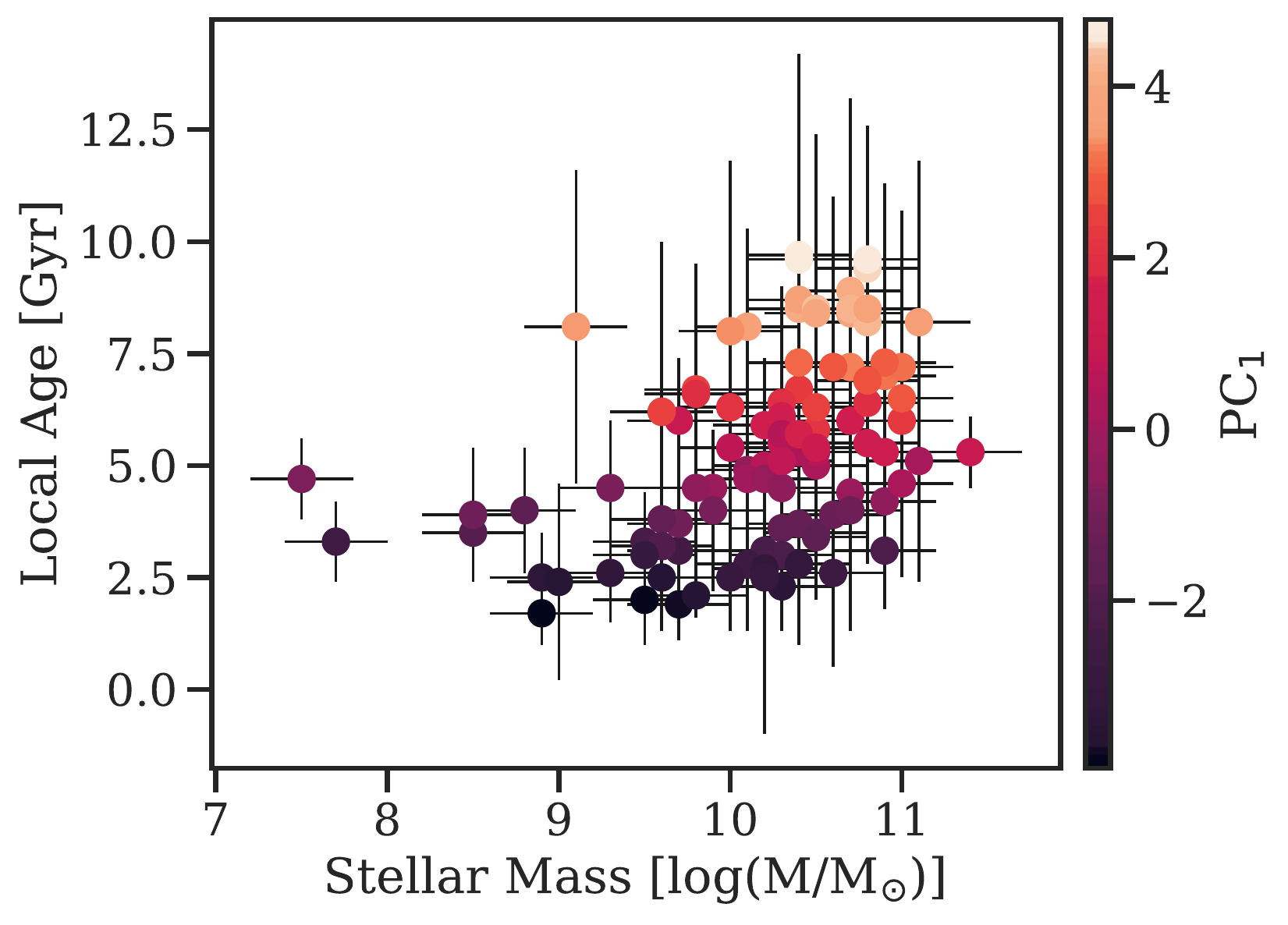}
    \caption{The distribution of \sn host galaxies analyzed in \citetalias{Rose2019} in the local age and stellar mass parameter space. These parameters are correlated, but not identical, especially with the wide range of ages possible in a high mass galaxy. The hosts are colored by their first principal component, a combination of \sn light curve stretch, galaxy stellar mass, and average stellar age. 
    All parameters are taken from \citetalias{Rose2019}.
    }
    \label{fig:mass-age}
\end{figure}

In \cref{fig:mass-age}, we show the data set from \citetalias{Rose2019} in the host galaxy local age-stellar mass plane. These two parameters are highly correlated, but not identical, especially since a large host galaxy may have a complicated star-formation history and a range of local stellar ages. The points are colored by the \citetalias{Rose2019} first principal component ($\mathrm{PC}_1 = 0.56x_1 - 0.10c - 0.54m_* - 0.63a_{l}$). PC$_1$ is a function of \sn light curve stretch, and color ($x_1$, $c$), along with rescaled host stellar mass and local age parameters ($m_*$, $a_l$). A detailed description of this component, and the resulting $4.7\sigma$ correlation with \hr{s}, can be found in section 7.4 of \citetalias{Rose2019}.
Unlike \hr{s}, \xo, or \co, PC$_1$ is smoothly distributed in the age-mass parameter space.

Most of the input data for \unity come from the original data release \citep{Campbell2013}. We use the redshifts relative to the cosmic microwave background radiation, the full \salt co-variance matrix, as well as the bias-corrected distance moduli.
The host galaxy properties come from Table 7 of \citetalias{Rose2019}. 
For this work, we remove the sample mean of the host galaxy properties in order to mimic the null mean of \sn light-curve parameters. We define this mean subtracted $\log$ of the stellar mass as $m_*$, and the mean subtracted local and global ages as $a_l$ and $a_g$ respectively.

\subsection{The Model}

From the generalized Tripp-like equation (\ref{eqn:standardization}) we can construct a specific standardization equation by expanding the summation over the host galaxy properties for our data set --- host stellar mass ($m_*$), average local ($a_l$) and global ages ($a_g$). Like all BHM, \unity calculates correlations against the ``true'' noiseless physical parameters, resulting in:
\begin{equation}
    \label{eqn:unity}
    \begin{split}
    \mu = m_B^{\mathrm{true}} - (&M_B + \alpha \xo^{\mathrm{true}} + \beta \co^{\mathrm{true}} + \gamma_m m_*^{\mathrm{true}} + \\&\gamma_{al} a_l^{\mathrm{true}} + \gamma_{ag} a_g^{\mathrm{true}} ) + \mathcal{N}(0, \su)~~.
    \end{split}
\end{equation}
The units for $\gamma_m$, $\gamma_{al}$, and $\gamma_{ag}$ are $\mathrm{mag}/\log_{10}(\mathrm{M}/\mathrm{M}_{\odot})$, $\mathrm{mag}/\mathrm{Gyr}$, and $\mathrm{mag}/\mathrm{Gyr}$ respectively.
We add an additional normally distributed scatter with a width of \su, $\mathcal{N}(0, \su)$, in order to account for variations in the physical properties that are not captured by the model \citep{Kelly2007}.
Note that \unity assumes that the distribution of each noiseless physical parameter used in the standardization equation can be represented as a Gaussian \citep{Gull1989}. This matches the \xo population well. For the \sn color, it will only accurately get the first two moments of the distribution but miss the expected skewness, an issue that is not significant for a fully linear analysis like this one, as demonstrated in \citet{Rubin2015}. However, a wide Gaussian parent population is not the true distributing of the host galaxy properties \citep[][\citetalias{Rose2019}]{Childress2014}. As a result, it will produce a slight prior against extreme values of stellar mass and age.

Each truth parameter ($i^{\mathrm{true}}$) is related to its observed value ($i^{\mathrm{obs}}$) via the addition of measurement noise ($\epsilon_i$):
\begin{equation}
    i^{\mathrm{obs}} =  i^{\mathrm{true}} + \epsilon_{i}~~.
\end{equation}
The measurement noise is unique per observation.
The associated analysis code for this paper can be found at \url{https://github.com/rubind/host_unity/tree/master/RRSG2020} with the analysis steps explained in the enclosed \texttt{makefile}.

\sn standardization can be improved by either an increase in precision or accuracy. A reduction of the post standardization scatter, beyond measurement and model uncertainties (\su), shows an increased precision of $M_B$ and as a result, the distance to any individual \sn. However, \su can easily be biased by poorly estimated uncertainties. In addition, a statistically significant new standardization term ($\gamma_i$) would allow for better constraints of redshift dependant systematics, improving \sn distance accuracy.

With this model, we are able to investigate the direct dependence of \sn peak absolute magnitude with host galaxy properties. The simultaneous standardization of both the \sn and host properties allows us to fully marginalize over the complex correlations between the parameters themselves and the standardization coefficients. This is the proper method to fit correlated standardization coefficients without bias \citep{Dixon2020}, rather than the serendipitous PCA investigation of \citetalias{Rose2019}.

\vspace{1em}
\section{Standardization Results}\label{sec:results}

\begin{deluxetable*}{l|CCCC|CCC}
\tablecolumns{8}
\tablewidth{0pt}
\tablecaption{Marginalized \sn Standardization Parameters\label{tab}}
\tablehead{
\colhead{Model} & \colhead{\su} & \colhead{\% change} & \colhead{$\alpha$} & \colhead{$\beta$} & \colhead{\gm} & \colhead{\gal} & \colhead{\gag}\\
\colhead{} & \colhead{[mag]} & \colhead{} & \colhead{[mag]} & \colhead{[mag]} & \colhead{[$\mathrm{mag}/\log_{10}(\mathrm{M}/\mathrm{M}_{\odot})$]} & \colhead{[mag/Gyr]} & \colhead{[mag/Gyr]}
}
\startdata
\salt & 0.122^{+ 0.019}_{-0.018} & \nodata & -0.15 \pm 0.017 & 3.1^{+0.3}_{-0.2} & \nodata & \nodata & \nodata \\ \hline
\salt \& \gm & 0.114 \pm 0.016 & 6\% & -0.17 \pm 0.018 & 3.2 \pm 0.2 & -0.11 \pm 0.03 & \nodata & \nodata \\
\salt \& \gal & 0.114 \pm 0.019 & 6\% & -0.20 \pm 0.03 & 3.2 \pm 0.3 & \nodata & -0.06^{+0.02}_{-0.03} & \nodata \\
\salt \& \gag & 0.113 \pm 0.017 & 7\% & -0.18 \pm 0.02 & 3.4 \pm 0.3 & \nodata & \nodata & -0.042 \pm 0.013 \\ \hline
\textbf{\salt,} $\boldsymbol{\gamma_{m}}$ \& $\boldsymbol{\gamma_{al}}$ & \boldsymbol{0.109 \pm 0.017} & \boldsymbol{10\%} & \boldsymbol{-0.19 \pm 0.03} & \boldsymbol{3.3 \pm 0.3} & \boldsymbol{-0.08^{+ 0.05}_{-0.04}} & \boldsymbol{-0.04 \pm 0.03} & \boldsymbol{\nodata} \\
\salt, \gm \& \gag & 0.113 \pm 0.016 & 7\% & -0.17 \pm 0.02 & 3.2 \pm 0.3 & -0.09^{+ 0.08}_{-0.09} & \nodata & -0.01^{+ 0.04}_{-0.03} \\
\salt, \gal \& \gag & 0.10^{+ 0.02}_{-0.03} & 10\% & -0.20^{+ 0.04}_{-0.08} & 3.3^{+ 0.4}_{-0.6} & \nodata & -0.06^{+0.14}_{-0.26} & -0.00^{+0.18}_{-0.10} \\ \hline
\salt, \gm, \gal \& \gag & 0.09^{+ 0.02}_{-0.04} & 25\% & -0.21^{+ 0.04}_{-0.08} & 3.1^{+ 0.5}_{-1.0} & -0.19^{+0.16}_{-0.26} & -0.13^{+0.13}_{-0.21} & -0.11^{+0.22}_{-0.13} \\ \hline
%
Only high-mass hosts & 0.115 \pm 0.018 & 6\% & -0.16 \pm 0.02 & 3.1 \pm 0.3 & \nodata & -0.03 \pm 0.02 & \nodata \\ 
Only low-mass hosts & 0.10 \pm 0.04 & 10\% & -0.22^{+0.06}_{-0.08} & 3.9^{+0.6}_{-0.7} & \nodata & -0.06^{+ 0.04}_{-0.05} & \nodata \\
\enddata
\tablecomments{Due to the order of magnitude smaller scale of $c$, $\beta$ is an order of magnitude larger than the other standardization parameters. The model in bold (\salt, \gm, \& \gal) improves both the precision and accuracy of \sn distances.
}
\end{deluxetable*}

First, we performed a standardization without any host galaxy parameters, as a null hypothesis. 
We obtain typical results, however the large alpha seen in the originally data release \citep[$\alpha = -0.22 \pm 0.02$,][]{Campbell2013} is no longer present. Instead, we calculated a more typical value of $-0.150\pm0.017$ \citep{Lampeitl2010,Marriner2011,Sako2014}. In addition, we calculated that for this data set \salt leaves an unexplained dispersion (\su) of $0.122 \pm 0.018 \un{mag}$.
A summary of the estimated model parameters for this and the other models explored in this paper are presented in \cref{tab}.

\subsection{Standardizing with Host Galaxy Properties}

We present our work by systematically building up to five standardization coefficients, allowing us to test and validate this methodology with the smaller sub-models.

\begin{figure}
    \centering
    \includegraphics[width=\columnwidth]{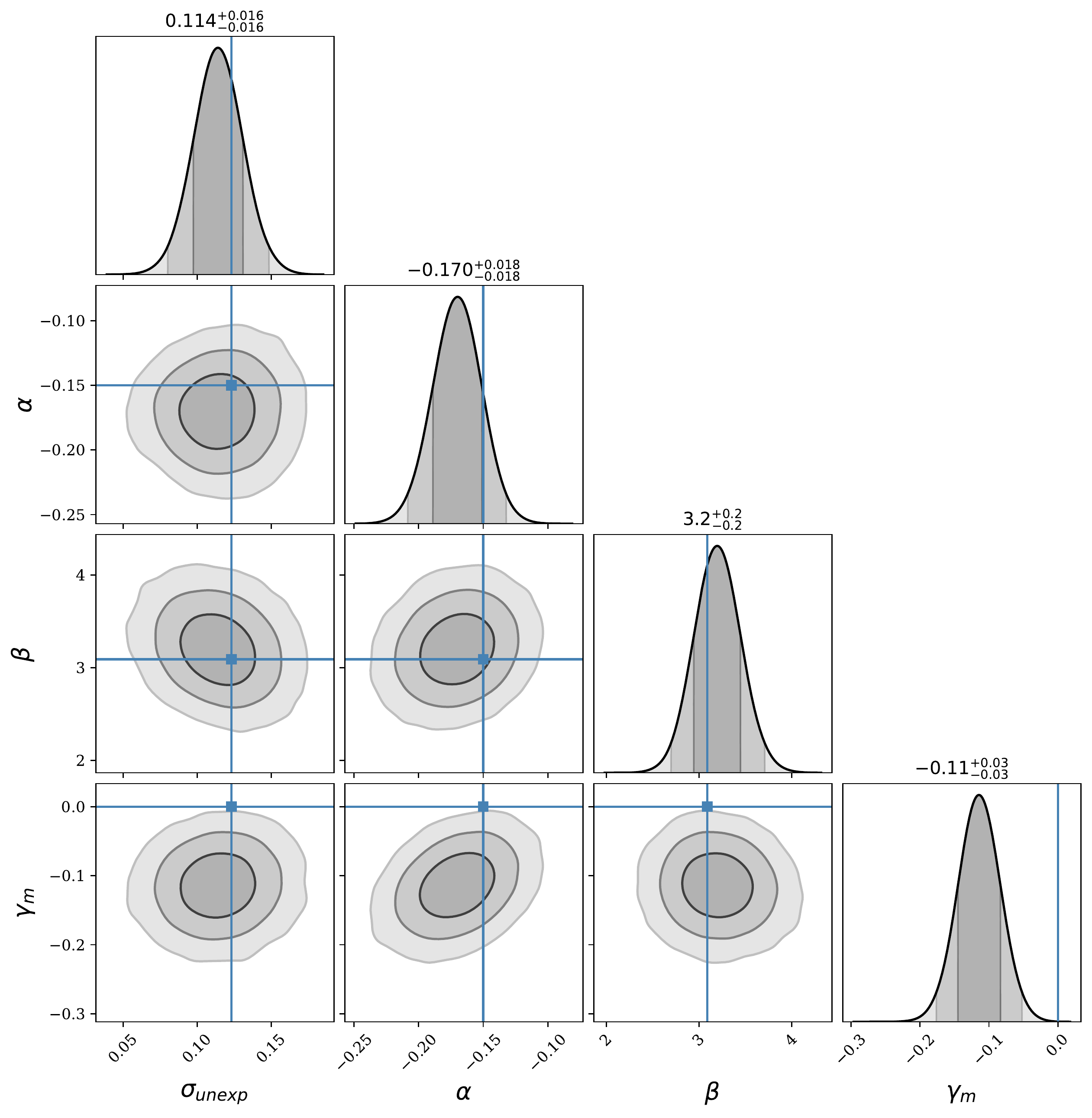}
    \caption{
    The posterior distribution, presented as a corner plot, of the \salt model parameters, including a linear standardization from host galaxy stellar mass ($\gamma_m$).
    Shaded regions show 1, 2, \& $3\sigma$ credible regions (dark, medium, and light respectively). Median and $1\sigma$ uncertainties are reported above each one dimensional marginalized distributions.
    Blue lines represent the median parameter values from the \salt only.
    The units for $\gamma_m$ are $\mathrm{mag}/\log_{10}(\mathrm{M}/\mathrm{M}_{\odot})$; the other variables all have units of $\un{mag}$.
    We find \gm to be significant ($3.7\sigma$) indicating a possible redshift dependant systematic if only \salt parameters are used. However, there is no improvement to \su with respect to a \salt only analysis.
    }
    \label{fig:oneHostParam}
\end{figure}

\begin{figure}
    \centering
    \includegraphics[width=\columnwidth]{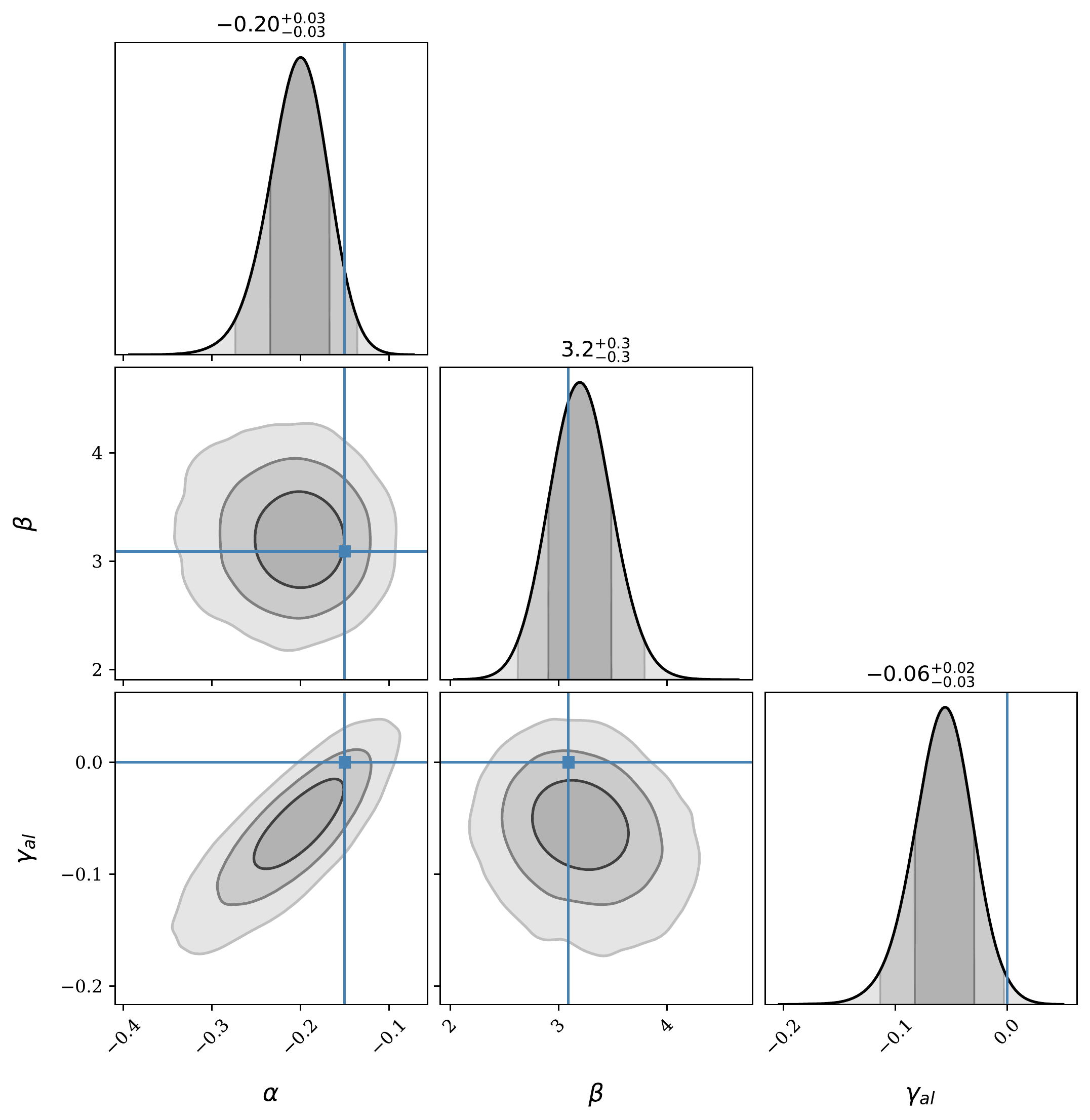}
    \caption{\sn standardization parameters for the \salt parameters and local stellar age, presented the same as \cref{fig:oneHostParam}.
    The standardization based on average localized stellar age ($\gamma_{al}$) has units of $\mathrm{mag}/\mathrm{Gyr}$.
    When simultaneously fitting, \gal is slightly more significant ($2.4\sigma$) than the $2\sigma$ seen from the sequential analysis of \citetalias{Rose2019}.
    The unexplained scatter ($\su = 0.114 \pm 0.019$) is not presented in this figure since it did not change from \cref{fig:oneHostParam}. In addition, standardizing local age meaningfully shifts $\alpha$ from the original $-0.15 \pm 0.017 \un{mag}$ to $-0.20 \pm 0.03 \un{mag}$.
    }
    \label{fig:local}
\end{figure}

\begin{figure}
    \centering
    \includegraphics[width=\columnwidth]{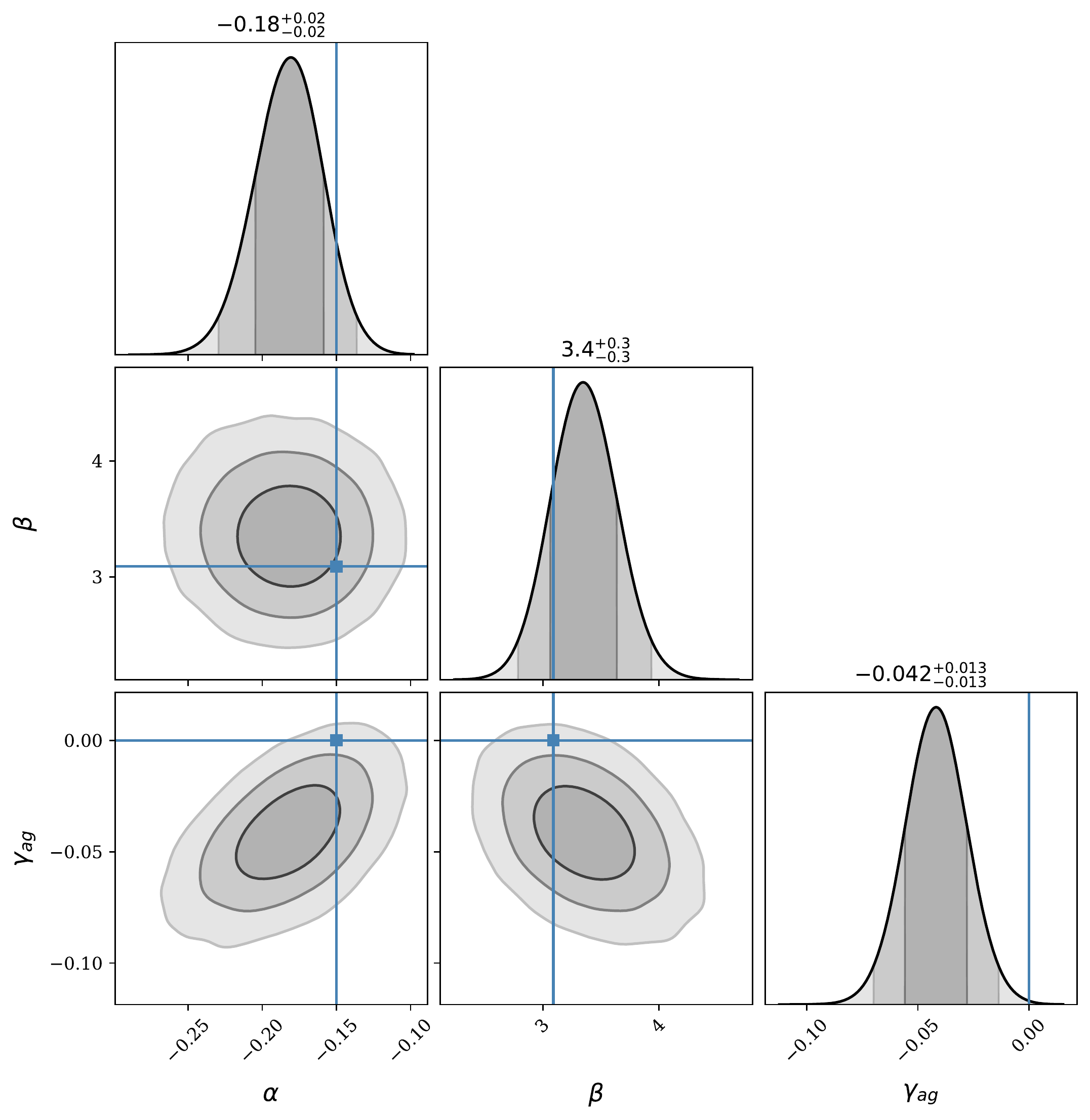}
    \caption{Same as \cref{fig:local} but standardizing with host integrated global stellar age (\gag).
    We see that \gag acts differently than \gal, contrary to the findings of \citetalias{Rose2019}.
    \gag is more significant ($3.2\sigma$) than \gal but also has a smaller effect. In addition, \gag is anti-correlated with $\beta$.
    The unexplained scatter ($\su = 0.113 \pm 0.017$) is not presented in this figure since it did not change from \cref{fig:oneHostParam}.
    }
    \label{fig:global}
\end{figure}

When adding only one host galaxy property at a time, each standardization coefficient is detected at $> 2 \sigma$ with the host stellar mass dependence seen at $3.7\sigma$ (\cref{fig:oneHostParam}). 
Though $\gamma_m$ is very significant \su did not significantly decrease ($< 0.01 \un{mag}$, a 6\% decrease). A host galaxy stellar mass correction would limit a possible redshift dependant bias, but does not increase the precision of any single \sn distance measurement. As was previously presented by \citet{Roman2018}, a correlation between $\alpha$ and $\gamma_i$ is present. This is especially true when standardizing on local stellar age (\gal) in \cref{fig:local}, where $\alpha$ strengthens from $-0.15 \pm 0.017 \un{mag}$ to $-0.20 \pm 0.03 \un{mag}$.
\Cref{fig:global} shows the posterior when standardizing with host galaxy global stellar age. None of these models reduce \su by more than 7\%.

Models with two host galaxy properties allow us to test if there is a statistical preference between the two properties, such as a possible preference between stellar mass and age. See \citet{Rigault2013}, \citet{Childress2014}, \citet{Kang2020}, and \citet{Rose2020b} for a brief history of this debate.

\begin{figure*}
    \centering
    \includegraphics[width=0.95\textwidth]{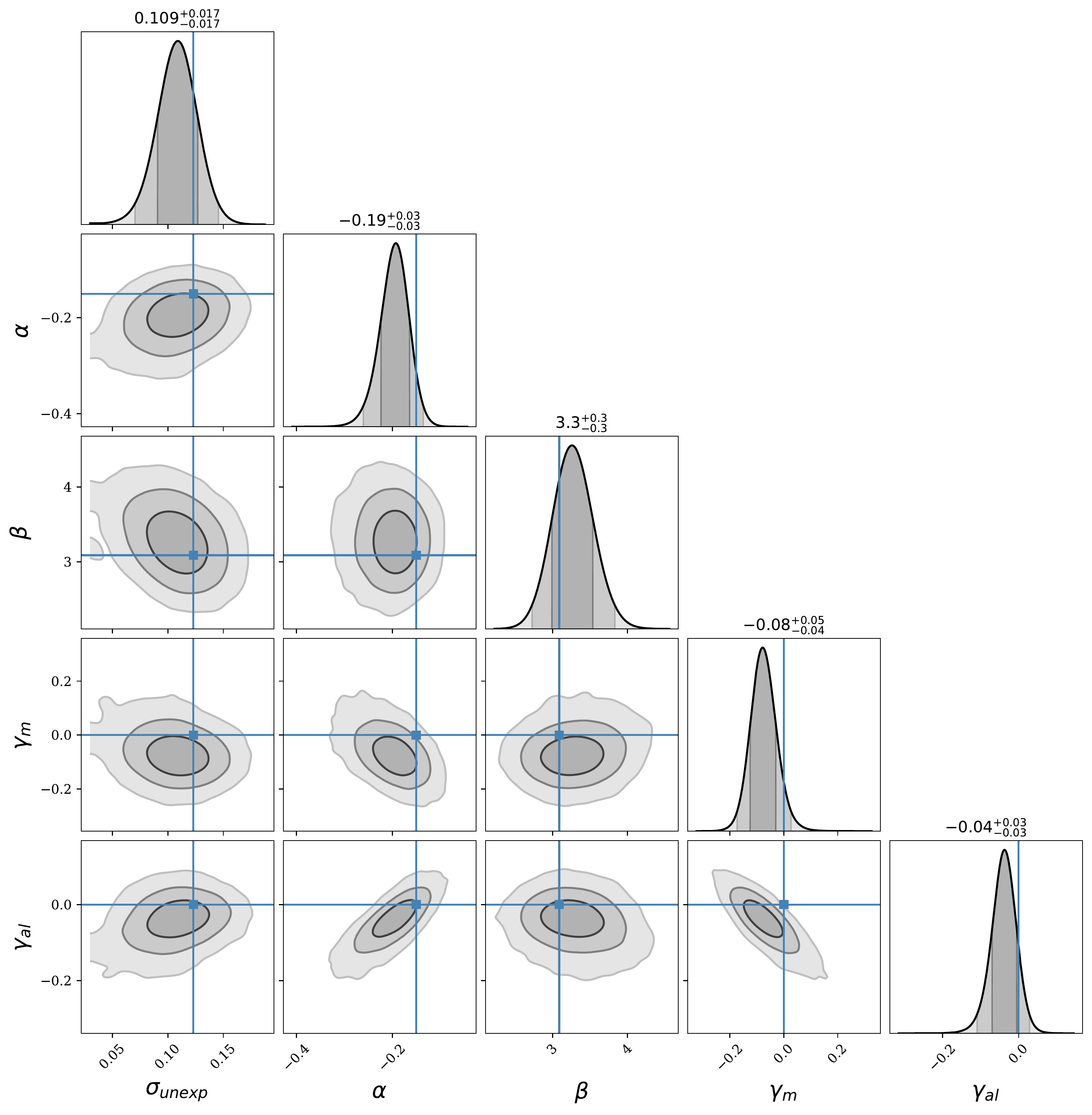}
    \caption{Same as \cref{fig:oneHostParam} but including local age ($\gamma_{al}$). 
    Each host galaxy parameter is less constrained individually ($<2\sigma$), however \su is reduced by $10\%$.
    While neither \gm or \gal are statistically non-zero when marginalized, their two dimensional credible region is non-zero at the $>3\sigma$ level indicating a possible redshift dependant bias. 
    In addition, $\alpha$ is highly correlated with host galaxy parameters, particularly \gal, and as a result has a $-0.04\un{mag}$ shift to its median value.
    The combination of \gm and \gal reduces \su and therefore increases the precision of each each \sn distance.
    }
    \label{fig:ml}
\end{figure*}

\Cref{fig:ml} shows a corner plot of the posterior when extending the \salt standardization methodology with both a stellar mass and local average stellar age term (\gm and \gal respectively). The $>2\sigma$ significance of \gm and \gal is no longer present. When marginalizing, they are only detected at $1.8\sigma$ and $1.3\sigma$, respectively. However, the $(\gm, \gal) = (0, 0)$ point is excluded at $>3\sigma$.
In addition, standardizing with both mass and local age reduces the unexplained scatter by 10\% ($\su=0.109\pm0.017$). Though the difference is smaller than the uncertainties, it is significantly more than when standardizing with only one host galaxy parameter and unlikely for a reanalysis of the same data set.

Finally, we standardize all three host galaxy terms. 
In this model, no host galaxy standardization term is significantly detected. However, there is a significant reduction in \su ($0.09 \pm 0.03\un{mag}$) and therefore an increase in precision of each \sn distance. 
Interestingly, the \gi-\su credible regions show that when any host galaxy standardization term approaches zero, \su increases.
We find --- as with all the sub-models --- that $\alpha$ is highly correlated with host galaxy parameters and as a result has a $-0.06\un{mag}$ shift to its median value.

\section{Discussion}\label{sec:discussion}

\subsection{Host effects on $\alpha$ and $\beta$}

There are several conclusions that can be drawn from the above analysis. First, the color coefficient ($\beta$) has minimal correlations with $\alpha$ and most host standardization properties (\gi). From the PCA of \citetalias{Rose2019}, \pc{1} contained no significant color component but \pc{2} was essentially only color. This means that at the parameter and the standardization coefficient levels, color is independent; a shift in another parameter does not significantly shift $\beta$. However, there is recent work that sees a relationship between the color parameter ($c$) and host galaxy standardization \citep{Brout2020,Gonzalez-Gaitan2020}.

Secondly, $\alpha$ does meaningfully change if you include host galaxy standardization. This can be seen in all of the $\alpha$--$\gamma_i$ contours. The extreme positive correlation between $\alpha$ and $\gamma_{al}$ is particularly evident in \cref{fig:local,fig:ml}. 
This correlation contradicts assumptions made in many previous host galaxy systematics studies (including \citetalias{Rose2019}) implying that they reported biased results. \citet{Dixon2020} presents a rigorous mathematical derivation of this bias.
Simultaneous fits of  $\alpha$ and $\gamma_i$ need to become standard practice. 
Since \citetalias{Rose2019} --- along with many others --- looked for correlation with \hr{}, they reported biased trends by using the wrong light-curve standardization parameters, in particular $\alpha$.

\subsection{Improved Standardization with Host Galaxy Properties}

Using the traditional \salt standardization, there is still $0.122^{+0.019}_{-0.018} \un{mag}$ of unexplained scatter. Using all three host parameters this is reduced to $0.09^{+0.02}_{-0.04} \un{mag}$, a $1.5 \sigma$ shift away from the original mean. The true statistical significance of this shift is larger since the two analyses were on the same data set.
Though using all three host galaxy parameters results in the smallest unexplained scatter, when standardizing with host stellar mass and average local age the unexplained scatter is nearly the same ($\su=0.109 \pm 0.017 \un{mag}$). In addition, the parameters of this smaller model are much better constrained. Interestingly, models with more significant host standardization parameters (i.e. stellar mass alone at $3.7\sigma$) do not always see the same large reduction of \su. Therefore, improvements in \sn distance accuracy and precision are not necessarily achieved simultaneously.
In terms of increased precision, a reduction in \su{} is seen every-time both stellar mass and local age are used, but not if only one is used.
This could be explained by over-estimated uncertainties. However, this particular trend implies $a_l$ would have overestimated uncertainties, something that would be unexpected since it is the hardest to measure.
Taken at face value, these results show that the local age is more important than the global age. This has been seen by \citet{Rigault2015,Rigault2018} but has not been confirmed by an outside group until now.

The anti-correlations seen in the $\gamma_i$-$\gamma_i$ posteriors are expected because of correlations in both measurement techniques and galaxy scaling relationships.
More interesting is \cref{fig:ml} where the $(\gm, \gal) = (0, 0)$ point is excluded at $>3\sigma$.
Were this point allowed, the anti-correlation would be trying to completely cancel each other. Anti-correlations with shifts away from (0, 0) in the second or fourth quadrant would indicate the preference, although slight, for one parameter over the other. A posterior maximum in the third (or the first) quadrant implies that they shift together. Knowing that the greatest reduction of \su is when two or more host parameters are used, implies that not only do they shift together, but they complement each other.

Standardizing with two ages, the (0, 0) point is not statistically excluded, though the uncertainties for this model are larger by nearly a factor of three. This is an example of what it looks like when \unity is splitting the standardization between two highly-correlated measurements. 
However, if one of the two host galaxy terms is stellar mass, then the (0, 0) point is excluded (e.g. \cref{fig:ml}).
\sne standardization is improved when it includes host stellar mass and local average stellar age; both terms are needed, as they are working together. 

\subsubsection{Metallicity over Explosion Mechanism}

The combination of stellar mass and age could point to the physics driving these effects. 
For example, a metallicity dominated systematic could show up as a combined effect of mass and age (the Mannucci relationship, \citealt{Mannucci2010}). However, a difference in peak luminosity from prompt and delayed explosions would have a stronger age effect than mass effect. This data supports the claim that \sn absolute magnitude correlates with a host galaxy property such as metallicity over a pure age effect like explosion ``channel.''

\subsubsection{Is it Dust?}

Instead of adding a host galaxy term ($\sum_{i=1}^N \gamma_i a_i$) to the Tripp-like standardization equation, \citet{Brout2020}, added a dust term ($(R_V+1) \times E(B-V)$, their Equation 13). They find that \rv values change drastically between low and high mass host galaxies.
We are able to test a few of their claims, with the caveat that the \sne in our analysis is a proper sub-sample of the \citet{Brout2020} data set.

Our above result, preferring a combination of mass and local age, does not refute the dust claims of \citet{Brout2020} since
this combination can indicate a shift in \rv. \citet{Salim2018} shows that attenuation changes with both mass and sSFR. 
At the high mass regime, there is an additional age-related spreading of the gradient in attenuation. If \rv is the main systematic cause, the dependence on age would be more pronounced in high mass hosts, a possible explanation for the results seen in \citet{Kang2020}.

\begin{figure}
    \centering
    \includegraphics[width=\columnwidth]{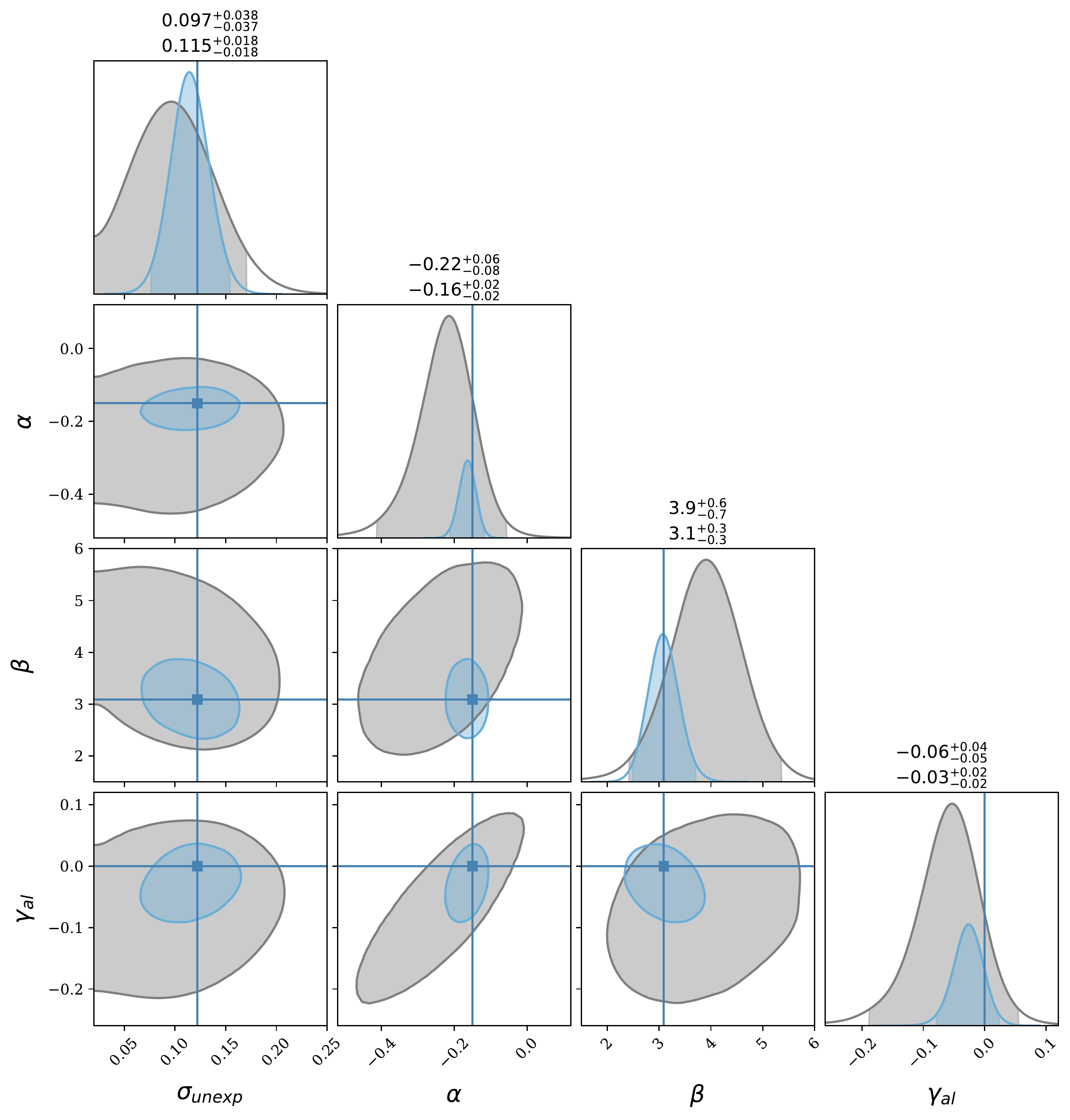}
    \caption{
    Same as \cref{fig:local} but splitting the data set based on host stellar mass and presented with $2\sigma$ credible regions.
    The gray shaded regions are for the low mass sub-sample ($< 10^{10} \un{M}_{\odot}$) and blue regions for the high mass sub-sample ($\geq 10^{10} \un{M}_{\odot}$). 
    The high mass sub-sample has a smaller age dependence, counter to the dust explanation, but they do have the lower $\beta$ ($\Delta\beta \approx 1 \un{mag}$) as seen in \citet{Sullivan2010}.
    However, the changing correlation of $\beta$ and $\gamma_{al}$ between the two sub-samples is Unforeseen.
    The only other model to see a correlation with $\beta$ is when standardizing with average global stellar age (\cref{fig:global}).
    Though dominated by uncertainties from the small samples, the complex interplay between these parameters indicates the need to reassess some of our assumptions, including the lack of cross terms in our standardization equations.
    }
    \label{fig:split}
\end{figure}

Following the model originally seen in \citet{Sullivan2010},
we split our sample into high ($\geq 10^{10} \un{M}_{\odot}$) and low ($< 10^{10} \un{M}_{\odot}$) stellar mass sub-samples ($N=72$ and $N=31$ respectively). In the high mass sub-sample, we look for both a larger age effect \citep{Salim2018} and a shift to lower $\beta$ \citep{Sullivan2010}.
Our results of this analysis can be seen in \cref{fig:split}.

We find that the high mass sub-sample has a smaller age dependence ($-0.027\pm0.02 \un{mag/Gyr}$ versus $-0.057\pm0.05\un{mag/Gyr}$), counter to the dust explanation. However, it does have the lower $\beta$ ($3.1\pm0.3 \un{mag}$ versus $3.9\pm0.6 \un{mag}$) as seen in \citet{Sullivan2010}.
Nonetheless, the key conclusion from splitting the sample on host galaxy stellar mass is that correlations with $\alpha$ and $\beta$ can unexpectedly change, i.e. the rotation between the sub-samples in the $\beta$--$\gamma_{al}$ plane. For the sample in its entirety (\cref{fig:local}) there was no correlation with $\beta$, however the high-mass sub-sample has a large $\beta$--$\gamma_{al}$ correlation. Interestingly, a correlation with $\beta$ is seen when standardizing with average global stellar age (\cref{fig:global}).
These sub-sample dependant correlations --- and the variations in these correlations --- likely indicate the need for a more complicated standardization method. Ultimately, our sub-samples are small and produce large uncertainties, making it difficult to understand exactly how these parameters are related.

When splitting on age (at the mean age of $5.2\un{Gyr}$) rather than stellar mass, we see a similar behavior as \cref{fig:split}, most notably a $\Delta\beta$ between the two samples of $\sim 0.7 \un{mag}$.

\subsection{Revisiting Rose et al. 2019}

This work presents an alternative analysis of the data from \citetalias{Rose2019}, allowing for the unique ability to compare these new results with the two major findings of the original analysis: a $2\sigma$ detection of a local or global age trend and a $4.7\sigma$ trend with a principal component mixing \sn light curve shape, host stellar age, and host stellar mass.

\subsubsection{Correlation with Stellar Age}

The dependence of \sn standardization on stellar ages seen in \citetalias{Rose2019} is smaller than what is found in this analysis.
\cref{fig:local,fig:global} show a $2.4\sigma$ and $3.2\sigma$ local and global age standardization respectively, contrary to the reduced significance from a simultaneous fit seen in \citet{Rose2020b}.
The work of \citet{Dixon2020} shows, mathematically, how a sequential fit (fitting $\alpha$ then \gi) with correlated variables produces a bias in the trend and its uncertainty. As we see here, the simultaneous fit of \unity produces a larger and more significant trend than originally seen in \citetalias{Rose2019}.

Unlike when fitting sequentially, simultaneous fitting finds local and global age to be unique.
Looking at the values, \gal is larger but also more than two times as uncertain as \gag. In addition, \gag has an anti-correlation with $\beta$ not seen in other models.
From this re-analysis, we find that \gal and \gag are not the same.

\subsubsection{Principal Component Analysis}

The PCA of \citetalias{Rose2019} found an eigenvector (\pcone) that strongly correlated with \hr{s}. Even a slight variation (see Figure 11b of \citetalias{Rose2019}) showed the possibility of an improved correlation.
\Cref{fig:ml} shows the results of standardizing with host stellar mass and local age.
Using Equations 10--12 along with Tables 8 and 9 of \citetalias{Rose2019}, we can convert the PC$_1$ values to the corresponding standardization parameters. This results in
\begin{equation}
    \mathrm{HR} = -0.028 x_1 - 0.063 c - 0.040 m_* - 0.015 a_l~~.
\end{equation}
The full conversion is presented in Appendix \ref{sec:pc1}.
Since this equation quantifies \hr{s} (HR), these standardization parameters are differences from the typical hostless standardization.
As expected PC$_1$ is an interesting direction in the parameter space, but it does not agree with the optimal standardization parameters seen in \cref{fig:ml}. When translated to a prescription of the HR, the optimized parameters are:
\begin{equation}\label{eqn:hr_result}
    \mathrm{HR} = 0.04 x_1 + 0.2 c + 0.08 m_* + 0.04 a_l~~.
\end{equation}{}

\subsection{Effect on \sn Cosmology}\label{sec:cosmo_effect}

The addition of standardizing \sn on stellar mass and average local stellar age is both statistically significant and increases the precision of each \sn distance. As a result this model, of all the models investigated in this work, best describes a possible bias of cosmological parameters. However, this data set is too small to constrain cosmological biases due to the fact that \gm and \gal are only detected at $\sim1\sigma$. Any comment would be unable to distinguish between no bias or the possibility of a large effect. 

With the correlation between $\alpha$ and \gal being nearly $-1$, you would expect that these two parameters would cancel each other out resulting in a similar cosmology with or without \gal.
Since \hr do not correlate with redshift, we can naively interpolate these changed standardization parameters, \cref{eqn:hr_result}, as a bias on cosmology by looking at how the standardization parameters mix with \sn population drift.

Between a redshift of zero and one, the average $x_1$ shifts by 0.5 \citep{Nicolas2020}. The work of \citet{Rubin2016} shows a larger shift, but this is in part because their low redshift sample is biased, by survey strategies, towards low $x_1$. 
\sn color, $c$, does not drift with redshift \citep{Rubin2016}.
The average \sn host galaxy mass shifts by $0.2 \un{dex}$ between redshift zero and one \citep{Sullivan2010,Strolger2020}.
Finally, there is no good estimate of the evolution of the local age, but it is likely small since it will be closely linked to the average delay time. On the other hand, the average global age shifts by $\sim 5 \un{Gyr}$ \citep{Childress2014}; but this number is dependant on the \sn delay time distribution.

Without a good estimate of the local age evolution we cannot use \cref{eqn:hr_result}, however we can use $\mathrm{HR} = (0.02 \pm 0.02) x_1 + (0.1 \pm 0.3) c + (0.08 \pm 0.08) m_* + (0.01 \pm 0.03) a_g$. Though global age did not standardize \sn as well as local age, it was similar enough to see the order of magnitude of a possible cosmological bias.
We find a change in distance from \xo drift of $0.01 \pm 0.01 \un{mag}$, from host galaxy stellar mass of $-0.03 \pm 0.03 \un{mag}$, and from global age of $-0.05 \pm 0.15 \un{mag}$. We estimate that a \salt only standardization would result in a bias of $-0.06 \pm 0.15 \un{mag}$ at redshift one.
Though the uncertainties are large, they are also underestimated since it does not account for the uncertainty in the amount of population drift.
Ultimately we need a larger data set to better constrains the standardization parameters and any cosmological parameter bias.

\section{Conclusions}\label{sec:conclusions}

Using a Bayesian hierarchical model, we are able to simultaneously fit the standardization parameters associated with \sn light curves, colors, and host galaxy properties. From this analysis, we are able to make six major conclusions on standardizing \sn with host galaxy properties.

\begin{enumerate}
    \item \Cref{fig:oneHostParam,fig:local,fig:ml} show that the \sn color standardization parameter, $\beta$, is not correlated with $\alpha$ or host galaxy standardization coefficients ($\gamma_i$).
    They also show that the light curve shape standardization parameter, $\alpha$, correlates with host galaxy standardization terms, requiring that these are simultaneously fit. This has previously been seen in \citet{Roman2018} and \citet{Rigault2018}.
    As a result, all sequential \hr versus host galaxy correlations are biased, including those of \citet{Rose2019}.
    
    \item However, \cref{fig:split}, shows that these correlations are not the same when splitting the data set on stellar mass. A correlation between \gal and $\beta$ developed for the high-mass sub-sample.
    These correlations and sub-sample dependencies point to a complex covariance that will require at least a simultaneous fit if not more complicated cross-terms.
    
    \item The statistical significance of any host galaxy correlation, $\gamma_i$, is an indicator of possible redshift dependent systematics but is not the same as improved \sn standardization precision, i.e. a reduction of \su. Standardizing on stellar mass alone produced the most significant host galaxy standardization coefficient ($3.7\sigma$) but the precision was improved with the addition of the average local stellar age, reducing the unexplained scatter on the same data by $\sim 1\sigma$ to $0.109 \pm 0.017 \un{mag}$.
    
    \item The one-dimensional marginalized significance of a parameter is not the whole story. Though neither host galaxy standardization term in the \salt plus \gm and \gal model is statistically significant individually, the two-dimensional marginalization shows a significant ($>3\sigma$) need for a modification to the \salt-based methodology.
    
    \item Local stellar age had a stronger impact than global stellar age, supporting the results of \citet{Rigault2015,Rigault2018}.
    
    \item A combination of mass and local stellar age is both statistically significant and improves the standardization precision, indicating a systematic from metallicity or dust rather than an age or explosion mechanism.
\end{enumerate}

We demonstrate, using a simultaneously fit, that the standardization coefficients are correlated in a non-trivial way that appears to be dependent on the stellar mass of the host galaxy.
Since neither stellar mass nor age are as significant or effective as a combination of them both, variations in line-of-sight dust or progenitor metallicity may be the physical source for the correlations between \sn and their host galaxies.

\acknowledgements
The authors would like to thank Rebekah Hounsell for encouragement and comments during the early stages of this research.
The authors would also like to thank Susana Deustua, Andrew Fruchter, Dan Scolnic, and David Jones for insightful discussions that informed several components of this paper.
Finally, we thank the anonymous referee for their time, attention, and clear report that improved this paper.
BR and DR acknowledge the support, in part, from NASA through grant NNG16PJ311I. BR also acknowledges support from NASA through grant NNG17PX03C.

\software{
click,
corner.py \citep{Foreman-Mackey2016}, 
emcee \citep{Foreman-Mackey2013}, 
kde\_corner,
Matplotlib \citep{matplotlib}, 
Numpy \citep{numpy}, 
Pandas \citep{pandas}, 
PyStan \citep{pystan},
Python, 
SciPy \citep{scipy}, 
Seaborn \citep{seaborn},
Stan \citep{Carpenter2017},
UNITY \citep{Rubin2015}
}
\vspace{1em} 

\appendix

\section{Conversion from PC$_1$ to a Change in Standardization Coefficients}\label{sec:pc1}
From \citet{Rose2019}, the principal component of interest (their Equation 11 and Table 9) is 
\begin{equation*}
    \text{PC}_1 = 0.557x'_1 - 0.103c' - 0.535m' - 0.627a' ~~.
\end{equation*}
This is for unit normal parameters ($i'$, defined in their Equation 10), and can be converted back to standard \sn parameters (plus a constant) by dividing by the standard deviations reported in their Table 8. Now we get
\begin{equation}
    \text{PC}_1 = 0.549x_1 - 1.24c - 0.775m/\log_{10}(\mathrm{M}/\mathrm{M}_{\odot}) - 0.297a/\mathrm{Gyr} ~~.
\end{equation}
In order to convert these to changes in standardization coefficients, we need to substitute $\text{PC}_1$ into the measured correlation with \hr{s} (HR), their Equation 12:
$$\text{HR} = 0.051 \un{mag} \times \text{PC}_1 - 0.012 \un{mag} ~~, $$
$$\text{HR} \propto (0.0280 \un{mag}) x_1 - (0.0632 \un{mag})c - (0.0395 ~\mathrm{mag}/\log_{10}(\mathrm{M}/\mathrm{M}_{\odot}))m - (0.0151 \un{mag/Gyr})a ~.$$
The last step, in order to compare these results with this paper, is to flip the sign on the coefficient in front of $x_1$ and reduce the coefficients to two significant digits:
\begin{equation}
    \text{HR} \propto (-0.028 \un{mag}) x_1 - (0.063 \un{mag})c - (0.040~ \mathrm{mag}/\log_{10}(\mathrm{M}/\mathrm{M}_{\odot}))m - (0.015 \un{mag/Gyr})a ~.
\end{equation}

\bibliographystyle{aasjournal}
\bibliography{library}


\end{document}